\documentclass[lettersize,journal]{IEEEtran}
\usepackage{amsmath,amssymb,amsfonts}%
\usepackage{array}
\usepackage[caption=false,font=normalsize,labelfont=sf,textfont=sf]{subfig}
\usepackage{textcomp}
\usepackage{stfloats}

\usepackage{url}
\usepackage{verbatim}
\usepackage{graphicx}
\usepackage{cite}
\usepackage{algorithmic}

\usepackage[utf8]{inputenc}
\usepackage[T1]{fontenc}

\usepackage{multirow}%
\usepackage{amsthm}%
\usepackage{mathrsfs}%
\usepackage{xcolor}%
\usepackage{textcomp}%
\usepackage{manyfoot}%
\usepackage{booktabs}%
\usepackage{algorithm}%
\usepackage{listings}%
\usepackage{hyperref}
\usepackage{caption}

\begin{document}

\title{Co-learning-aided Multi-modal-deep-learning Framework of Passive DOA Estimators for a Heterogeneous Hybrid Massive MIMO Receiver}

\author{Jiatong Bai, Feng Shu, Fuhui Zhou, Qinghe Zheng, Bo Xu, Baihua Shi, Yiwen Chen, Weibin Zhang, Xianpeng Wang 

\thanks{Corresponding author: Feng Shu}
\thanks{Jiatong Bai, Xianpeng Wang, Bo Xu and Yiwen Chen is with the School of Information and Communication Engineering, Hainan University, Haikou, 570228, China. (e-mail: 18419229733@163.com; wxpeng2016@hainanu.edu.cn; 996458@hainanu.edu.cn; cyw1978650281@163.com ).}
\thanks{Feng Shu is with the School of Information and Communication Engineering and Collaborative Innovation Center of Information Technology, Hainan University, Haikou 570228, China, and also with the School of Electronic and Optical Engineering, Nanjing University of Science and Technology, Nanjing 210094, China. (e-mail: shufeng0101@163.com).}
\thanks{Fuhui Zhou is with the College of Electronic and Information Engineering
Nanjing University of Aeronautics and Astronautics, Nanjing 210094, China. (e-mail: zhoufuhui@ieee.org).}
\thanks{Qinghe Zheng is with the School of Intelligent Engineering, Shandong Management University, Jinan 250357, China. (e-mail: zqh@sdmu.edu.cn).}
\thanks{Weibin Zhang and Baihua Shi is with the Nanjing University of Science and Technology, Jiangsu, Nanjing, 210094, China. (e-mail: weibin.zhang@njust.edu.cn; shibh56h@foxmail.com).}}




\maketitle
\begin{abstract}

Due to its excellent performance in rate and resolution, fully-digital (FD) massive multiple-input multiple-output (MIMO) antenna arrays has been widely applied in data transmission and direction of arrival (DOA) measurements, etc. But it confronts with two main challenges: high computational complexity and circuit cost. The two problems may be addressed well by hybrid analog-digital (HAD) structure. But there exists the problem of phase ambiguity for HAD, which leads to its low-efficiency or high-latency. Does exist there such a MIMO structure of owning low-cost, low-complexity and high time efficiency at the same time. To satisfy the three properties, a novel heterogeneous hybrid MIMO receiver structure of integrating FD and heterogeneous HAD ($\rm{H}^2$AD-FD) is proposed and corresponding multi-modal (MD)-learning framework is developed. The framework includes three major stages: 1) generate the candidate sets via root multiple signal classification (Root-MUSIC) or deep learning (DL); 2) infer the class of true solutions from candidate sets using machine learning (ML) methods; 3) fuse the two-part true solutions to achieve a better DOA estimation. The above process form two methods named MD-Root-MUSIC and MDDL. To improve DOA estimation accuracy and reduce the clustering complexity, a co-learning-aided MD framework is proposed to form two enhanced methods named CoMDDL and CoMD-RootMUSIC. Moreover, the Cramer-Rao lower bound (CRLB) for the proposed $\rm{H}^2$AD-FD structure is also derived. Experimental results demonstrate that our proposed four methods could approach the CRLB for signal-to-noise ratio (SNR) > 0 dB and the proposed CoMDDL and MDDL perform better than CoMD-RootMUSIC and MD-RootMUSIC, particularly in the extremely low SNR region. 

\end{abstract}


\section{Introduction}
Direction of arrival (DOA) estimation is pivotal in various domains, including target localization\cite{10285515}, radio
propagation measurement\cite{10458050}, and wireless communications\cite{6873240}. Its integration with massive multiple input multiple output (MIMO) techniques yields ultra-high-accuracy the desirable signal direction for numerous wireless communication techniques\cite{dong2022low,8333706,9530396} such as beamforming and directional modulation. Additionally, integrated sensing and communication is emerging as a burgeoning research domain, DOA estimation based on massive MIMO plays a core role within ISAC, applicable in realms\cite{8839973,10176899,10287137}, like 6G, human-machine interaction, and the Internet of Things.

The existing DOA estimation methods can be roughly categorized into 
the classical spatial spectral-based techniques, parametric-based estimation methods, the sparsity sensing-based techniques and deep learning (DL)-based methods. 
To the best of our knowledge, the most applied techniques are subspace-based conventional DOA measurement method. 
The multiple signal classification (MUSIC) estimation method in \cite{schmidt1982signal} partitioned the total column space of the covariance matrix 
into signal and noise subspaces
, thus achieving accurate source localization. However, MUSIC algorithms were computationally burdensome due to the fact that spatial spectral peak searches. To address this issue, 
the rotational invariance technique (ESPRIT) method for signal parameters estimation was introduced \cite{roy1986esprit} to directly obtain DOA estimation from closed-form solutions. 
The MUSIC and ESPRIT algorithms obtained the subspace by the array covariance matrix' eigenvalue decomposition, leading to a significant increase in computational complexity for large-scale antenna array.
The root multiple signal classification (Root-MUSIC) algorithm \cite{barabell1983improving,rao1989performance} transformed the spectral search into a problem of finding high-order polynomials roots, significantly enhancing computational efficiency. Based on above classical algorithms, several improved algorithms\cite{zhang2009multi,pesavento2000unitary,yan2014real,zhang2017improved} had been proposed to achieve superior DOA estimation.
Furthermore, motivated by the potent feature extraction and precise prediction capabilities of deep learning, several scholars had investigated the DL-based DOA measurement methods\cite{8845653,papageorgiou2021deep, ma2024deep}. For instance, DOA estimation performance was improved in low signal-to-noise ratio (SNR) by convolutional neural networks (CNN) in \cite{papageorgiou2021deep}.

Massive MIMO-based DOA estimation systems can obtain ultra-high angular resolution and estimation accuracy, but a sharp increase in computational complexity and circuit cost also occurs.
The hybrid analog and digital (HAD) structure converted signals from multi antennas into baseband signals through a single analog-to-digital converter (ADC) and radio frequency (RF) chain, thereby efficiently improving the utilization of RF chains and ADCs while reducing circuit costs\cite{zhang2016angle,chuang2015high,9684752}.
However, due to the fact that the HAD structure treats each sub-array as a virtual antenna, the fact that the spacing between virtual antennas is equal to the number subarray antennas multiplied by the half-wavelength leads to phase ambiguity and generates pseudo-solutions.
In \cite{shuHADDOA2018tcom}, a low-complexity HAD structure and three maximum received power-based methods were proposed to achieve high-precision and multi-time-slot DOA estimation, where the Root-MUSIC-HDAPA algorithm realized a hybrid Cramér-Rao Lower Bound (CRLB) with lower complexity. However, this method required $M+1$ ($M$ represented the number of subarray antennas) time-slots to eliminate phase ambiguity.
To minimize estimation delay, the enhanced fast FHAD-Root-MUSIC algorithm in \cite{11111} achieved two time-slots phase ambiguity elimination by partitioning multi subgroups and performing analog beamforming for each subgroup.
Additionally, in \cite{chen2022two} and \cite{DBLP:journals/corr/abs-2201-04452}, rapid DOA estimation algorithms were proposed under four different HAD sub-connection structures to achieve single time-slot phase ambiguity elimination with a large performance loss.

Based on the above analysis, the DOA estimation method using HAD sub-connection structures consists of two steps: firstly, a set of candidate solution sets was obtained by a conventional methods like Root-MUISC. Then the pseudo-solutions in the candidate solution sets were eliminated to find the true DOA estimation value. In summary, the existing DOA estimation methods with phase ambiguity elimination capability generally require at least two time-slots, leading to low time-efficiency issues. Whereas, this paper will explore the MDL framework of DOA measurement in the heterogeneous hybrid massive MIMO system, focusing on eliminating phase ambiguity with high time-efficiency, high-performance, low cost and complexity. The major contributions of our research are outlined as follows: 
\begin{enumerate}
    \item To address the issue of the high latency or low time-efficiency of DOA estimation of the conventional HAD, a novel heterogeneous hybrid MIMO receiver structure, called $\rm{H}^2$AD-FD, of integrating a FD subarray with a $\rm{H}^2$AD structure is developed. The $\rm{H}^2$AD part comprises $H$ groups, where different groups have the equal number of subarrays but varying amounts of antennas. Also, within each group, every subarray maintains an identical number of antennas. In the co-learning scenario, the FD subarray will provide a good starting sample point of true solution class and will accelerate the process of the following clustering. Moreover, the corresponding CRLB for proposed $\rm{H}^2$AD-FD structure is also derived.
    \item  Based on the above structure, a multi-modal-learning framework is proposed to achieve a low-latency DOA estimation. Its primary stages are as follows: 1) all sub-array groups in $\rm{H}^2$AD produce the set of candidate solutions using Root-MUSIC and FD subarry estimate the DOA value using Root-MUSIC or CNN-based methods, where CNN is designed to be a five-layer neural network; 2) infer the class of true solutions from candidate sets using ML methods; 3) fuse the two-part solutions to output an final DOA estimation. The above process forms two methods for DOA measurement without ambiguity as follows:  MD-Root-MUSIC and MDDL. The corresponding fusion coefficients are computed by CRLB. Simulation results indicate that our proposed MDDL performs better than MD-Root-MUSIC. 
    
    \item To further enhance the DOA estimation accuracy, a co-learning-aided multi-modal (Co-MD) framework is proposed, Compared with the previous MD framework,  a relation between FD and $\rm{H}^2$AD modals is established. The FD modal will provide a prior knowledge to the $\rm{H}^2$AD modal. This relation will not only improve performance but also reduce the clustering complexity. Similar to the the previous MD framework, the two new corresponding methods are as follows:  CoMDDL and CoMD-RootMUSIC, where MDDL is short for MD deep learning (DL).
    Simulation results illustrate that our proposed CoMDDL and MDDL outperform CoMD-RootMUSIC and MD-RooMUSIC, respectively.

\end{enumerate}

The remainder of this research is organized as follows. The system
model of an developed heterogeneous MIMO structure with FD and
$\rm{H}^2$AD is described in Section \ref{sec_sys}. In Section \ref{sec_proposed}, a MD-learning (MDL) for DOA estimator
based on FD$\rm{H}^2$AD structure is proposed. In Section \ref{sec_CNN}, 
the clustering and fusion methods for our proposed frameworks are proposed. Moreover, section \ref{sec_perf} analyzes the performance of the proposal structure and approaches. Section \ref{sec_simu} presents the experimental results, with conclusions provided in Section \ref{sec_con}

\emph{Notations:} In this paper, uppercase letters and lowercase
letters in bold typeface (i.e., $\mathbf{A}$, $\mathbf{a}$) denote matrices and vectors, respectively. Signs $\|\cdot\|$, $|\cdot|$, $(\cdot)^H$, $(\cdot)^T$, $\Re\{\cdot\}$, $\Im \{\cdot\}$, $\angle\{\cdot\}$ and $\mathbf{diag}(\cdot)$ denote norm, modulus, conjugate transpose, transpose, real part operations, imaginary part operations, phase component and diagonal, respectively. Besides, $\mathbb{C}[\cdot]$ and $\mathbb{E}[\cdot]$ denote the complex-valued matrix and the expectation operator, respectively. $\mathbf{Tr}(\cdot)$ is the matrix trace.

\section{System model}\label{sec_sys}

\begin{figure}[!htb]
	\centering
	\includegraphics[width=3.5in]{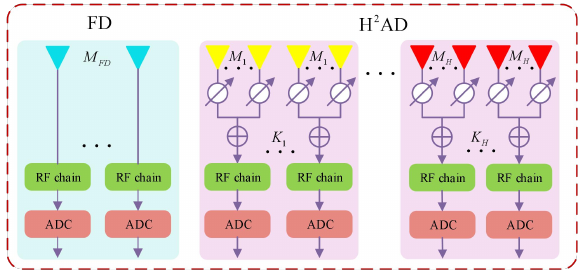}\label{fig: sub_figure1}
	\caption{Proposed heterogeneous hybird array with FD and $\rm{H}^2$AD, where $\rm{H}^2$AD has $H$ groups and group $h$ has $K_h$ subarrays with each subarray having $M_h$ antennas, $h\in\{1,2,\cdots,H\}$ }
	\label{fig_sys_mod}
\end{figure}

Figure~\ref{fig_sys_mod} presents the proposed heterogeneous hybrid structure, which involves a fully-digital (FD) and a $\rm{H}^2$AD array. A far-field narrowband signal $x(t)e^{j2\pi f_ct}$, where $x(t)$ and $f_c$ are the baseband signal and the carrier frequency, respectively. 
Considering the FD with $M$ antennas, the output signal is represented as
\begin{align} \label{11}
	\mathbf{y}_{FD}(t)=\mathbf{a}_{FD}(\theta_0)x(t)+\mathbf{w}(t),
\end{align}
where $\mathbf{w}(t)\sim\mathcal{C}\mathcal{N}(0,\sigma^2_w\mathbf{I}m)$
is the additive white Gaussian
noise (AWGN) vector, $\mathbf{a}_{FD}$ denotes the array manifold vector expressed as
\begin{align}
	\mathbf{a}_{FD}\left(\theta_0\right)=\left[1, e^{j \frac{2 \pi}{\lambda} d \sin \theta_0}, \cdots, e^{j \frac{2 \pi}{\lambda}(M-1) d \sin \theta_0}\right]^T,
\end{align}

The $\rm{H}^2$AD array is divided into $H$ groups, group $h$ has $K_h$ subarrays with each subarray containing $M_h$ antennas, i.e.,
\begin{align}
M =\sum_{h=1}^{H}N_h=\sum_{h=1}^{H}K_hM_h,
\end{align}
In this heterogeneous array, $M_1\neq M_2 \neq\cdots\neq M_H$, and each group is a homogeneous structure. It is particularly noted that it is preferred to choose the values of $M_1, M_2,\cdots, M_H$ to be primes.

Considering $\psi_{h,k,m}$ represents the phase corresponding to analog beamforming
, then the output of the $k$th subarray is
\begin{align}
	y_{h,k}(t)=\frac{1}{\sqrt{M_h}}\sum_{m=1}^{M_h}x(t)e^{j2\pi f_c(t-\tau_{h,k,m})-j\psi_{h,k,m}}+w_{h,k}(t)
\end{align}
where $w_{h,k}(t)\sim\mathcal{C}\mathcal{N}(0,\sigma^2_w)$ is the AWGN vector.
$\tau_{h,k,m}$ are the propagation delays established by the direction of the signal source with respect to the array expressed as
\begin{align}
	\tau_{h,k, m}=\tau_0-\frac{(km-1)d \sin \theta_0}{c},
\end{align}
where $\tau_0$ denotes the propagation delay from the radiating source to the reference point of the antenna array, and $c$ is the speed of light.

Stacking the outputs $y_{h,k}(t)$ of all $K_h$ subarray and the recieved signal vector of $h$th group is defined as 
\begin{align} \label{yq}
\mathbf{y}_h(t)=\mathbf{\Upsilon}_{A,h}^H\mathbf{a}_h(\theta_0)x(t)+\mathbf{w}(t),
\end{align}
where $\mathbf{w}(t)=\left[w_1(t), w_2(t), \ldots, w_{K_h}(t)\right]^T\in\mathbb{C}^{K_h \times 1}$ is the AWGN vector, $\mathbf{a}_h(\theta_0)\in\mathbb{C}^{N_h \times 1}$ is the array manifold vector expressed as
\begin{align}
	\mathbf{a}_h\left(\theta_0\right)=\left[1, e^{j \frac{2 \pi}{\lambda} d \sin \theta_0}, \cdots, e^{j \frac{2 \pi}{\lambda}(N_h-1) d \sin \theta_0}\right]^T,
\end{align}
and $\mathbf{\Upsilon}_{A,h}$ denotes a block diagonal matrix
\begin{align}
{\mathbf{\Upsilon _{A,h}}} = \left[ {\begin{array}{*{20}{c}}
{{\gamma _{A,h,1}}}&0& \cdots &0\\
0&{{\gamma _{A,h,2}}}& \cdots &0\\
 \vdots & \vdots & \ddots & \vdots \\
0&0& \cdots &{{\gamma _{A,h,K}}}
\end{array}} \right],
\end{align}
where $\mathbf{\gamma}_{A,h,k}$ is the $k$-th block diagonal element defined by
\begin{align}
	\mathbf{\gamma}_{A,h,k}=\frac{1}{\sqrt{M_h}}\left[e^{j\psi_{h,k,1}},e^{j\psi_{h,k,2}},\cdots,e^{j\psi_{h,k,M_h}}\right]^T,
\end{align}

Passing through analog-to-digital converter (ADC), the (\ref{yq}) becomes
\begin{align}	\mathbf{y}_h(n)=\mathbf{\Upsilon}_{A,h}^H\mathbf{a}_h(\theta_0)x(n)+\mathbf{w}(n),
\end{align}
where $n=1,2,\cdots,L$, $L$ is the number of snapshots.

\section{Proposed Multi-modal-learning frameworks for DOA estimators based on heterogeneous hybrid structure}\label{sec_proposed}

In this section, a MDL for DOA estimator based on heterogeneous hybrid structure are proposed to accelerate the elimination of pseudo-solutions, form a class of true solutions, and make a fusion of all true solutions to achieve an enhancement on DOA estimation.
\subsection{Proposed MDL frameworks }
\begin{figure}[!http]
	\centerline{\includegraphics[width=3.4in]{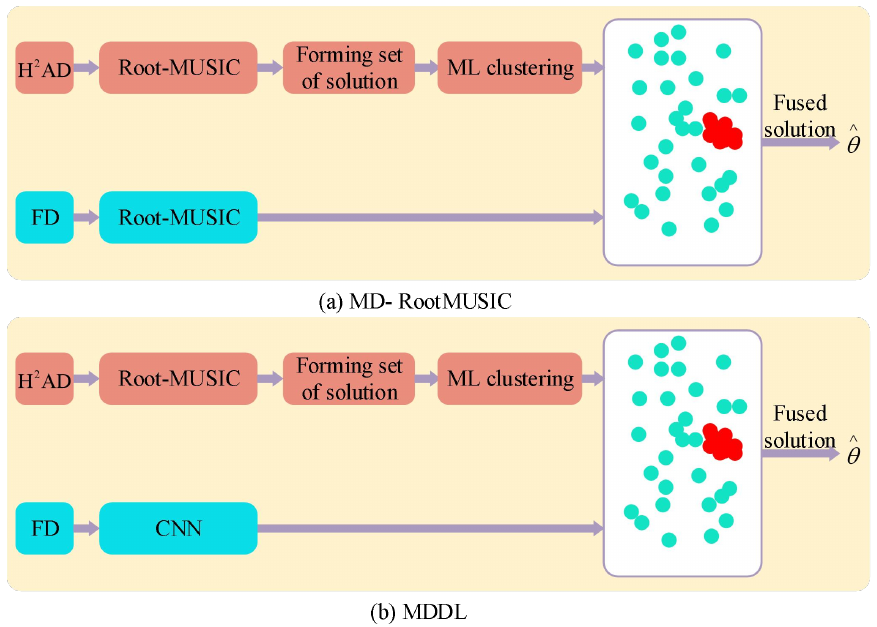}}
	\caption{Multi-modal framework for estimating DOA.\label{fig_alg_flow1}}
\end{figure}

\begin{figure}[!http]
	\centerline{\includegraphics[width=3.4in]{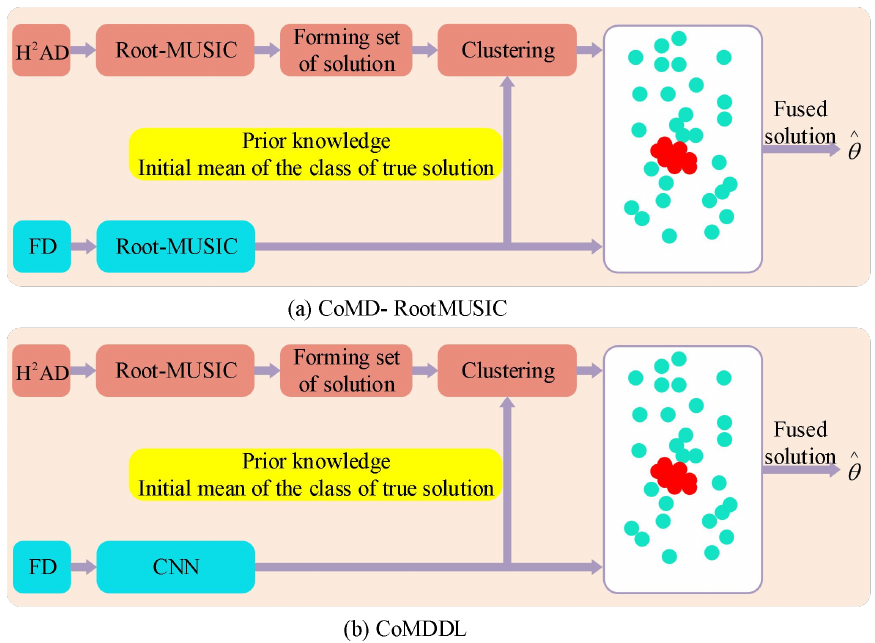}}
	\caption{Co-learning-aided multi-modal framework for estimating DOA.\label{fig_alg_flow2}}
\end{figure}

As shown in Figure \ref{fig_alg_flow1}, the proposed MDL framework consists of three main steps: 1) all sub-array groups in $\rm{H}^2$AD form the set of candidate solutions using Root-MUSIC and FD subarray directly generates the coarse estimated value of true solution using Root-MUSIC or CNN-based method.
2) infer the class of true solutions from candidate sets of $\rm{H}^2$AD using ML methods. 
3) fuse the two-part solutions to output an enhanced DOA estimation.
The above process forms two methods for DOA estimation named MD-Root-MUSIC and MDDL, shown in Figure \ref{fig_alg_flow1}(a) and (b), respectively.
Then, a Co-MDL framework is proposed, shown in Figure \ref{fig_alg_flow2}, to further enhance the DOA estimation accuracy. Compared with the MDL framework, a relation between FD and $\rm{H}^2$AD modals is established by utilizing the FD subarray to provide a better starting sample point of the true solution class. The two new corresponding methods is CoMDDL and CoMD-RootMUSIC, shown in Figure \ref{fig_alg_flow2}(a) and (b), respectively.

\subsection{Candidate solutions of $\rm{H}^2$AD}
Based on the above analysis, each subarray can be treated as a virtual antenna in group $h$th. Assume that the analog beamforming vector $\mathbf{\gamma}_{A,h,k}=\frac{1}{{\sqrt {{M_h}} }}{\left[ {1, \cdots ,1} \right]^T}$, the output vector of $K_h$ subarrays is
\begin{align} \label{yq1}
	\mathbf{y}_h(n)&=[y_1(n),y_2(n),\cdots,y_{K_h}(n)]^T\\\nonumber
	&=\frac{1}{{\sqrt {{M_h}} }}\mathbf{a}_{M_h}(\theta_0)b_h(\theta_0)x(n)+\mathbf{w}(n),
\end{align} 
where $\mathbf{w}(n)=[w_1(n),\cdots,w_{K_h}(n)]^T$, and $\mathbf{a}_{M_h}$ 
is regarded as the array manifold per sub-array, which can be formed by
\begin{align} \label{}
	\mathbf{a}_{M_h}\left(\theta_0\right)=\left[1, e^{j \frac{2 \pi}{\lambda} M_h d \sin \theta_0}, \cdots, e^{j \frac{2 \pi}{\lambda}(K_h-1) M_h d \sin \theta_0}\right]^T,
\end{align}
and $b_h(\theta_0)$ is defined as
\begin{align} \label{}
	b_h\left(\theta_0\right)  =\sum_{m=1}^{M_h} e^{j \frac{2 \pi}{\lambda}(m-1) d \sin \theta_0} 
	=\frac{1-e^{j \frac{2 \pi}{\lambda} M_h d \sin \theta_0}}{1-e^{j \frac{2 \pi}{\lambda} d \sin \theta_0}}
\end{align}
The covariance matrix of the output vector $\mathbf{y_h(n)}$ is
\begin{equation}
\begin{aligned} \label{}
	\mathbf{R}_{h} & =\mathbb{E}\left[\mathbf{y_h}(n) \mathbf{y_h}(n)^H\right] \\
 &=b_h(\theta_0)\mathbf{a}_{M_h}(\theta_0)\mathbf{R}_{ss}(b_h(\theta_0)\mathbf{a}_{M_h}^H\left(\theta_0\right)+\mathbf{R}_{ww}\\
	& =\frac{1}{M_h} \sigma_s^2\left\|b_h\left(\theta_0\right)\right\|^2 \mathbf{a}_{M_h}\left(\theta_0\right) \mathbf{a}_{M_h}^H\left(\theta_0\right)+\sigma_w^2 \mathbf{I},
\end{aligned}
\end{equation}
where $\sigma_s^2$ is the SNR of the receive signal and $\sigma_s^2=\mathbf{R}_{ss}=\mathbb{E}\left[{\left| {s\left( n \right)} \right|^2}\right]$. Besides, the eigenvalue decomposition (EVD) of $\mathbf{R}_{h}$ is performed as
\begin{align} \label{}
	\mathbf{R}_{h} =[\mathbf{U}_S \mathbf{U}_N]\Sigma[\mathbf{U}_S \mathbf{U}_N]^H,
\end{align}
where $\mathbf{U}_N$ is ${K_h} \times({K_h} - 1)$ noise subspace, $\mathbf{U}_S$ is ${K_h} \times 1$ signal subspace. $\Sigma=diag(\sigma_s^2+\sigma_w^2,\sigma_w^2,\cdots,\sigma_w^2)$ is the $K_h\times K_h$ diagonal matrix.

Then, to compute so-called ``MUSIC spectrum'' of the virtual antenna array
\begin{align} \label{}
	P_{\rm{H}^2AD}(\theta)=\frac{1}{\left\|b_h(\theta)\right\|^2\left\|\mathbf{a}_{M_h}^H(\theta)\mathbf{U}_N\mathbf{U}_N^H\mathbf{a}_{M_h}\right\|},
\end{align}

The direction corresponding to the maximum value of $P_{\rm{H}^2AD}(\theta)$ is the final DOA estimation. Furthermore, this paper applies the Root-MUSIC \cite{1993The} to obtain the optimal emitter direction owing to its remarkable asymptotic solution and low complexity. 

The polynomial equation can be defined as

\begin{equation}\label{p}
\begin{aligned} 	f(\theta)&=b_h^H(\theta)\mathbf{a}_{M_h}^H(\theta)\mathbf{U}_N\mathbf{U}_N^H\mathbf{a}_{M_h}(\theta)b_h(\theta)\triangleq f(z)\\
	&=\frac{2-z^{-1}-z}{2-z^{-\frac{1}{M_h}}-z^{\frac{1}{M_h}}}\sum_{i=1}^{K_h}\sum_{j=1}^{K_h}z^{-(i-1)}\mathbf{Q}_{ij}z^{j-1}\\
   &\triangleq f(\varphi)=0,
\end{aligned}
\end{equation}
where $\mathbf{Q}=\mathbf{U}_N\mathbf{U}_N^H$ and $\mathbf{Q}_{ij}$ represents the element located in the $i$th row and $j$th column of matrix $\mathbf{Q}$.

\begin{align} \label{z}
	z=e^{j\varphi_h},
\end{align}
and
\begin{align} \label{}
	\varphi_h=\frac{2 \pi}{\lambda} M_h d \sin \theta_0,
\end{align}

The above equation (\ref{p}) has $2K_h-2$ roots, that is, ${Z}_{R M}=\left\{{z}_i, i \in\{1,2, \cdots, 2 K_h-2\}\right\}$. Then, using equation (\ref{z}) to obtain the set of DOA estimates

\begin{align} \label{}
	\hat{\Theta}_{R M}=\left\{\hat{\theta}_i, i \in\{1,2, \cdots, 2 K_h-2\}\right\},
\end{align}
where
\begin{align} \label{}
	\hat{\theta}_i=\arcsin \left(\frac{\lambda \arg z_i}{2 \pi M_h d}\right)
\end{align}
Then, the DOA estimation $\hat\theta_{h}$ of $h$th array group can select the nearest root to the unit circle

\begin{align} \label{}
	\hat{\varphi}_h=\frac{2\pi}{\lambda} M_hdsin\hat{\theta_{h}}
\end{align}

Observing the function $f(\varphi) $ has a period of $2\pi$ about $\varphi$, thus the presence of phase ambiguity that requires elimination,
\begin{align} \label{}
	f(\hat{\varphi}_h)=f(\hat{\varphi}_h+2\pi j)
\end{align}
which the $h$-th group yields feasible solutions including $M_q$ values as follows
\begin{align} \label{THETA_q}
	\hat{\Theta}_{h}=\left\{\hat{\theta}_{h, j_h}, j_h \in\{1,2, \cdots, M_h\}\right\}
\end{align}
where
\begin{align} \label{}
	\hat{\theta}_{h, j_h}=\arcsin \left(\frac{\lambda\left(\arg (e^{j\hat{\varphi}_h})+2 \pi j\right)}{2 \pi M_h d}\right) .
\end{align}
Combing all $Q$ groups gives
\begin{align}\label{angleCandAll}
	\begin{split}
		\left \{
		\begin{array}{ll}
			\frac{2\pi}{\lambda}M_1dsin\theta_{1,j_1}=\hat{\varphi}_1+2\pi j\\
			\frac{2\pi}{\lambda}M_2dsin\theta_{2,j_2}=\hat{\varphi}_2+2\pi j\\
			\quad\quad\quad\vdots\quad\quad\quad\quad\quad\quad\quad\vdots\\
			\frac{2\pi}{\lambda}M_Hdsin\theta_{H,j_H}=\hat{\varphi}_H+2\pi j
		\end{array}
		\right.
	\end{split}
\end{align}
where $j_h\in\{1,2,\cdots,M_h\}$. Furthermore, forming the candidate solution set expressed as
\begin{align} \label{}
	\hat{\Theta}=\left\{\hat{\Theta}_{1}, \hat{\Theta}_{2} \cdots, \hat{\Theta}_{H}\right\}
\end{align}

Based on the above discussion, the total candidate set $\hat{\Theta}$ includes $\sum\limits_{h = 1}^H {{M_{_h}}}$ solutions, where each candidate set ${\hat{\Theta}_h}$ has a true angle and $M_{_h}-1$ pseudo-solutions, thus finding the true angle is a challenging task. For the $h$th group, the true angle and pseudo solutions can be represented as follows
\begin{align} \label{qqqqq}
	\hat{\theta}_{t,h}=\theta_0+\varepsilon_h
\end{align}	
\begin{align} \label{cc}
\hat{\theta}_{l,h,m}=\theta_0+\varepsilon_h+\Gamma_{h,m},~m=1,2,\cdots,M_h-1,
\end{align}
where $\varepsilon_h$ is the estimated error and $\Gamma_{h,m}$ is a constant decided by $M_h$ and $m$. When there is no noise, $\varepsilon_h\rightarrow 0$, the intersection of all candidate angle sets, there is only one value, $\theta_0$. The true and false solutions can be formed by
\begin{align} \label{bb}
	\left\{ \begin{array}{l}
{\hat{\theta}_{t,1}} \approx {\hat{\theta}_{t,2}} \approx  \cdots  \approx {\hat{\theta}_{t,Q}} \approx {\theta _0}\\
{\hat{\theta}_{l,1,m}} \ne {\hat{\theta}_{l,2,m}} \ne  \cdots  \ne {\hat{\theta}_{l,Q,m}} \ne {\theta _0}
\end{array} \right.
\end{align}
Via equations (\ref{cc}) and (\ref{bb}), pseudo solutions are unequal to each other even noise $\varepsilon_h\rightarrow 0$, that is $\hat{\theta}_{l,h,m}=\theta_0+\Gamma_{h,m}$, due to the fact that $K_H$ sun-array groups contains different number of antennas. Also, the distance between true angles tends to zero. This means the distances among true solutions is far smaller than the distances among pseudo solutions. Hence, in what follows, the minimum distance will be used as a metric of clustering.

Therefore, the set of true solutions can be defined as
\begin{equation}\label{eeeee}
	\hat{\Theta}_{t}=\left\{ \hat{\theta}_{t,1},\hat{\theta}_{t,2},\cdots,\hat{\theta}_{t,H} \right\}
\end{equation}
where $\hat{\theta}_{t,h}$ is the inferred true angle of the $h$ array group.


\subsection{RootMUSIC DOA estimation of FD subarray}
The FD subarray estimates the initial coarse DOA value via state-of-the-art methods like Root-MUSIC and deep learning. The covariance matrix and EVD of equation (\ref{11}) are expressed as 
\begin{equation}
\begin{aligned} \label{}
	\mathbf{R}_{FD} & =\sigma_s^2\mathbf{a}_{FD}(\theta_0)\mathbf{a}_{FD}^H(\theta_0)+\mathbf{I}\\
 & =[\mathbf{E}_S \mathbf{E}_N]\Sigma[\mathbf{E}_S \mathbf{E}_N]^H,
\end{aligned}
\end{equation}
where $\sigma_s^2$ denotes the SNR of the receive signal and
$\Sigma=diag\left\{ {\left[ {\sigma _s^2,1, \cdots ,1} \right]} \right\}$.

Then the corresponding spatial spectrum function is obtained as follows 
\begin{align} \label{}
	P_{FD}(\theta)=\frac{1}{\left\|\mathbf{a}_{FD}^H(\theta)\mathbf{E}_N\right\|^2},
\end{align}

Constructing the $2(M-1)$ polynomial equation through Root-MUSIC, and thus the nearest root to the unit circle is DOA estimation $\hat{\theta}_{FD}$, that is, initial mean of the class of true solution.
\begin{align} \label{FD}
	\hat{\theta}_{FD}=\arcsin \left(\frac{\lambda}{2 \pi d} \arg z_{FD}\right)
\end{align}

\subsection{Proposed CNN-based DOA measurement for FD subarray}
This section proposes a CNN model to form initial coarse DOA value for FD subarray, providing a better starting sample point of
true solution class compared to RootMUSIC. Specifically, the feature is extracted by the convolution layer and then the DOA estimation is inferred by the fully connected layer (FC).

\subsubsection{Datasets and its Label}
Considering that the DL model mainly involves real-valued computations, the final preprocessing procedure of the input data is to convert them to real description while retaining the imaginary part. Thus, the CNN model' input data $\mathbf{I}\in {\mathbb{R}^{M \times M \times 3}}$ is a real-valued matrix, expressed as

\begin{equation}
\mathbf{I} = \left\{ \begin{array}{l}
{\mathbf{I}_{:,:,1}} = \Re \{ {\mathbf{R}_{FD}}\} \\
{\mathbf{I}_{:,:,2}} = \Im \{ {\mathbf{R}_{FD}}\} \\
{\mathbf{I}_{:,:,3}} = \angle \{ {\mathbf{R}_{FD}}\} 
\end{array} \right.
\end{equation}

Assuming $S$ denotes the number of training sets, input samples of CNN can be defined as $\mathbf{I'}=\{ \mathbf{I}_1, \mathbf{I}_2, \cdots, \mathbf{I}_S\}$. The angle of the training data is chosen from the $p$ grid points of $[-\theta,\theta]$, where $\theta\in[1^\circ,90^\circ]$ and $p$ is interval cell. Thus, the desired grid is defined as $\Delta =\{-\theta,\cdots,-p,0^\circ,p,\cdots,\theta\}$ and the binary vectors corresponding to the training angles are taken as labels. For example, $\Delta =\{-45^\circ,\cdots,-1^\circ,0^\circ,1^\circ,\cdots,45^\circ\}$ having $\left| {\Delta} \right|=91$ grid points when $\theta=45^\circ$ and $p=1^\circ$. Besides, the angle $-45^\circ$ is converted to a binary vector $\nu_1 =[1,0,\cdots,0]^T\in \mathbb{R}^{91\times1}$. Based on the above analysis, the training dataset of CNN is

\begin{equation}\label{train}
\mathbf{I'}=\{ \mathbf{I}_1, \mathbf{I}_2, \cdots, \mathbf{I}_S\}
\end{equation}
And the label set corresponding to the training dataset is 
\begin{equation}\label{label}
\mathbf{V}=\{\nu_1,\nu_2,\cdots,\nu_S\}
\end{equation}

Therefore, the training sets is denoted by 
\begin{equation}
  {\mathbf{X}}=\{ \mathbf{I'},\mathbf{V}\}=\{ (\mathbf{I}_1,\nu_1), (\mathbf{I}_2,\nu_2), \cdots, (\mathbf{I}_S,\nu_S)\},  
\end{equation}
where $(\mathbf{I}_s,\nu_s)$ is the $s$-th training sample.

Moreover, the training phase utilizes the covariance matrix, while the testing process adopts the sample covariance matrix defined by
\begin{equation}
\overline{\mathbf{R}}_{FD}= \frac{1}{L}\sum\limits_{l = 1}^L {\mathbf{y}_{FD}(t)} \mathbf{y}_{_{FD}}^H(t)
\end{equation}

\subsubsection{CNN Structure}
\begin{figure}[!htb]
	\centering	\includegraphics[width=3.4in]{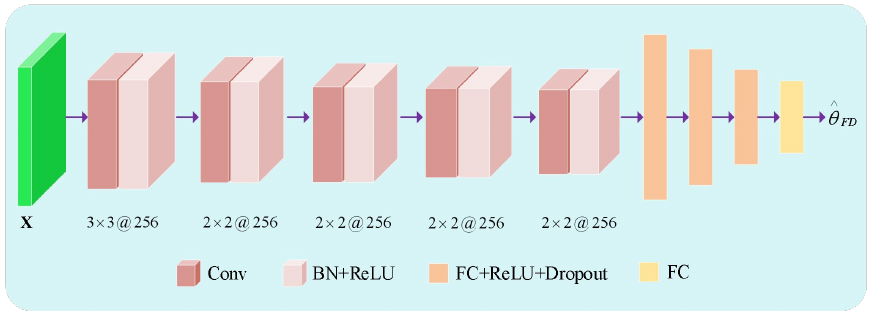}
	\caption{The poposed CNN--based DOA estimation for FD Array.}\label{CNN1}
\end{figure}

In addition to the impact of training dataset, the performance of CNN largely depends on the network structure. 
The objective of CNN-based DOA estimation is $f: {\mathbb{R}^{M \times M \times 3}}\to \hat{\theta}_{FD}$.
We introduce a CNN network including 2D convolutional layers and FC layers, illustrated in Figure.\ref{CNN1}, to predict DOA estimation for FD Array. To this end, initial mean of the class of true solution is generated to accelerate the inference of true solution sets for $\rm{H}^2$AD Array. The mapping of input to output by proposed CNN can be defined as

\begin{equation}
\nu  = f\left( \mathbf{I} \right) = {f_{{F_4}}}\left( { \cdots {f_{{F_1}}}\left( {{f_{{C_5}}}\left( { \cdots {f_{{C_1}}}\left( \mathbf{I} \right)} \right)} \right)} \right)
\end{equation}
where $f_{{C_i}} (i=1,\cdots,5$) is $i$-th convolutional layer, including a 2D convolutional layer with $256$ filters, a batch normalization (BN) layer and a ReLU layer. And ReLU can be denoted as

\begin{equation}
{\rm{ReLU}}\left( x \right) = \max \left( {0,x} \right)
\end{equation}
Then, expanding the output of $f_{{C_5}}$ to a vector via a flatten layer. $f_{{F_j}} (j=1,\cdots,4$) is the $j$-th FC layer, involving a dense layer, a ReLU layer and a dropout layer. The dropout layer can inhibit over-fitting and enhance the regularization properties during the training phase of the CNN.

\subsubsection{Loss Function}
The training datasets and the set of label in Eqs. (\ref{train}) and (\ref{label}) are used to train the CNN model. Furthermore, we add the $L_2$ regularization term to mitigate the over-fitting and improve the model generalization. Therefore, the loss function is defined as

\begin{equation}
L\left( {{{\hat \theta }_{FD}}} \right) = \frac{1}{2}\left\| {\overline \nu  \left( {{{\hat \theta }_{FD}}} \right)} \right\|_{^2}^2+\mu {{\left\| W \right\|}^{2}},
\end{equation}
\begin{equation}
\overline \nu  \left( {{{\hat \theta }_{FD}}} \right) = \nu \left( {{{\hat \theta }_{FD}}} \right) - \mathop \nu \limits^ \wedge  \left( {{{\hat \theta }_{FD}}} \right),
\end{equation}
where $\mathop \nu \limits^ \wedge  \left( {{{\hat \theta }_{FD}}} \right)$ and $\nu \left( {{{\hat \theta }_{FD}}} \right)$ represent the true label and the predicted value, respectively. $\mu$ denotes the weight attenuation coefficient, set to $0.0005$. $W$ is the weighting parameter of the FC layer. 
The loss function $L$ is minimized to train the proposed model.

\subsubsection{Training Details}
\begin{figure}[!htb]
	\centering	\includegraphics[width=3.0in]{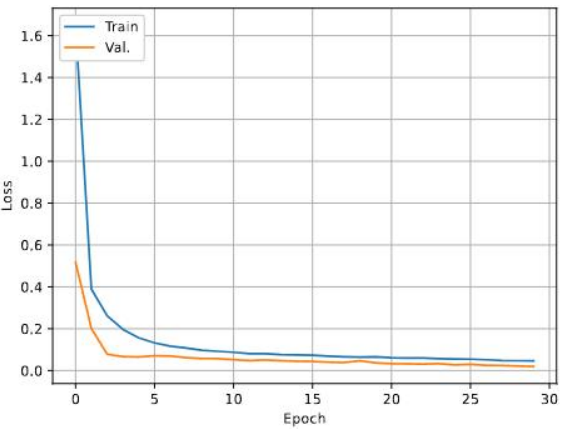}
	\caption{The training and validation loss of the CNN.}\label{PROCESS}
\end{figure}

The Adam optimizer is used to train the proposed CNN network with a learning rate of $0.0001$, which can avoid overfitting the optimal solution. Training batch\_size is set to $32$ and epochs is $30$ in the entire operation of training.The $L_2$ regularization coefficient is $0.0005$. Dropout layers with rate $0.3$ in the FC layers. Signal impinging directions are $\Delta =\{-90^\circ,\cdots,-1^\circ,0^\circ,1^\circ,\cdots,90^\circ\}$ and $\Delta =\{-45^\circ,\cdots,-1^\circ,0^\circ,1^\circ,\cdots,45^\circ\}$ in training and testing processes, respectively. Figure.\ref{PROCESS} shows the training and validation loss of the CNN model without over-fitting occurring.

\section{Proposed clustering and fusion methods for our proposed frameworks
}\label{sec_CNN}
In this section, we will focus on how to infer the class of true solutions from the candidate sets and fuse two-part solutions to form an enhanced DOA estimation. To make a high-accuracy classification, several high-performance methods: CoMDDL, CoMD-RootMUSIC, MDDL and MD-RooMUSIC are depicted as follows.

\subsection{Improved Hierarchical Clustering-aided MDDL and MD-RooMUSIC}
\begin{figure}
	\centerline{\includegraphics[width=3.5in]{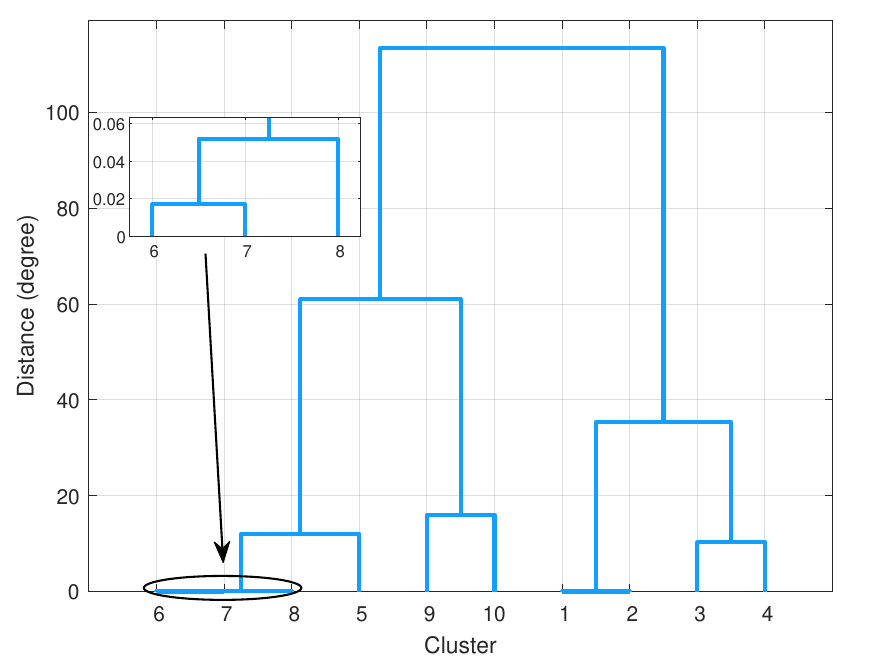}}
	\caption{The implementation of the hierarchical clustering for candidate angle set\label{fig_hir_cluster}}
\end{figure}
Due to the fact that true angles of all group arrays in $\rm{H}^2AD$ will be close to each other,
the extraction of true angles could be transferred into a cluster problem.

The hierarchical clustering is able to calculate the distance between different clusters and generate a hierarchy of clusters. There are two strategies in hierarchical clustering: agglomerative and divisive. Agglomerative, so called "bottom-up" approach, is more suitable in this application. This method starts by treating all $H$ points as $H$ clusters. Then, calculate all distances between two different clusters and merge the two closest clusters into a new cluster. After that, repeat the above step until only one cluster is left. 

In our problem, all $H$ point is 1-D points and we are trying to find five nearest points. Thus, Euclidean distance is selected and the mean value of a cluster is chosen for computing distance between clusters.
As shown in Figure \ref{fig_hir_cluster}, we draw a dendrogram for the proposed heterogeneous HAD structure with $M_1=2$, $M_2=3$ and $M_3=5$. The $3$ true solutions are cluster 6, 7 and 8, respectively. It is obvious that the true solutions were merged into one cluster with a very low distance. Thus, the method could be terminated when the first cluster having $H$ is born. Then, selected points could also be  combined by (\ref{thetaEstConbine}) (\ref{wqFinal1}) and (\ref{wqFinal2}).
The whole method can be called as Improved-HS and shown in Algorithm \ref{alg:hire}.
\begin{algorithm}[t]
	\caption{Improved-HS-aided MD.}\label{alg:hire}
	\begin{algorithmic}
		\STATE {\textbf{Input:}}$~\mathbf{y}(n),~ n=1,2,\cdots,L.$
		\STATE \hspace{0.5cm} \textbf{Initialization:} ~divide $\mathbf{y}(n)$ into $\mathbf{y}_{FD}(n)$ and $\mathbf{y}_h(n)$, $h=1,2,\cdots,H$. Calculate the $CRLB_{h}$ and $CRLB_{FD}$.
		\STATE \hspace{0.5cm} \textbf{for} $h=1,2,\cdots,H$ \textbf{do},
		\STATE \hspace{1cm} perform the root-MUSIC method for $\mathbf{y}_h(n)$, and obtain the solution set, $\hat{\Theta}_{h}$.
		\STATE \hspace{0.5cm} \textbf{end for}
		\STATE \hspace{0.5cm} treat $H$ solutions as $H$ clusters
		\STATE \hspace{0.5cm} \textbf{repeat}
            \STATE \hspace{0.5cm} Calculate the $\hat{\theta}_{FD}$ by the root-MUSIC method for $\mathbf{y}_{FD}(n)$.
		\STATE \hspace{1cm} Compute the distance between different clusters and combine the nearest two clusters as a new cluster
		\STATE \hspace{0.5cm} \textbf{Until} there is one cluster having $H$ solutions
		\STATE \hspace{0.5cm} The cluster having $H$ solutions is regarded as $\hat{\Theta}_{t}$
            \STATE \hspace{0.5cm} Integrate the $\hat{\Theta}_{t}$,(\ref{wqFinal1}) and (\ref{wqFinal2}) into (\ref{thetaEstConbine}) to get $\hat{\theta}$.
		\STATE {\textbf{Output:}} $\hat{\theta}$
	\end{algorithmic}
\end{algorithm}

\subsection{Proposed CoMDDL and CoMD-RootMUSIC methods}
Based on the proposed CoMD framework in Figure \ref{fig_alg_flow2}, to eliminate pseudo-solutions for each group by transferring the estimation value $\hat{\theta}_{FD}$ of the FD sunarray into the feasible solution sets $\hat{\Theta}_{h}$ from the $h$-th group of $\rm{H}^2$AD part, which accelerates the true optimal angle of the $h$-th group.
\begin{align} \label{gsProb}	\hat{\theta}_{\rm{H}^2AD,h}=\underset{\hat{\theta}_{h,j_h}\in\hat{\Theta}_h}{\arg \min }\left\|\hat{\theta}_{FD}-\hat{\theta}_{h,j_h}\right\|^2
\end{align}
Based on (\ref{qqqqq})-(\ref{eeeee}), the set of true solutions is
\begin{equation}
	\hat{\Theta}_{t}'=\left\{ \hat{\theta}_{\rm{H}^2AD,1},\hat{\theta}_{\rm{H}^2AD,2},\cdots,\hat{\theta}_{\rm{H}^2AD,H} \right\}
\end{equation}
By fusing the two-part true solution to obtain the optimal DOA estimate
\begin{align} \label{thetaEstConbine}
	\hat{\theta}= w_{FD}\hat{\theta}_{FD} + \sum_{h=1}^{H} w_{h}\hat{\theta}_{\rm{H}^2AD,h} 
\end{align}

The mean square error (MSE) of the $\hat{\theta}$ is
\begin{equation} 
\begin{aligned} \label{MAEA}
	&\mathbf{MSE}(\hat{\theta})=\mathbb{E}\left[\left(\hat{\theta}-\theta_0\right)^2\right]\\
	&=\sum_{h=1}^{H} w_{h}^2
 \mathbb{E}\left[\left(\hat{\theta}_{\rm{H}^2AD,h}-\theta_0\right)^2\right]+w_{FD}^2\mathbb{E}\left[\left(\hat{\theta}_{FD}-\theta_0\right)^2\right]\\
 &=\sum_{h=1}^{H} w_{h}^2\mathbf{MSE}(\hat{\theta}_{\rm{H}^2AD},h)+w_{FD}^2\mathbf{MSE}(\hat{\theta}_{FD})\\
 &\ge \sum_{h=1}^{H} w_{h}^2CRLB_{h}+w_{FD}^2 CRLB_{FD},
\end{aligned}
\end{equation}
Thus
\begin{align} \label{w_op}
	&\min_{w_h,w_{FD}}~~~\sum_{h=1}^{H} w_{h}^2CRLB_{h}+w_{FD}^2 CRLB_{FD}\nonumber\\
	&s.t. ~~~~~~~~~w_{FD} + \sum_{h=1}^{H} w_{h}=1,
\end{align}

According to \cite{T.Engin-09} and \cite{shuHADDOA2018tcom}, the $C R L B_{FD}$ and is given by
\begin{align}
C R L B_{FD} = \frac{{{\lambda ^2}}}{{8L{\pi ^2}\mathbf{SNR}{{\cos }^2}{\theta _0}\mathop {{d^2}}\limits^ - }}
\end{align} 

Referring to \cite{shuHADDOA2018tcom}, the Fisher information matrix of the
$h$th $\rm{H}^2$AD group array can be expressed by (\ref{FIMHAD11}), where
\begin{figure*}[t]
	\begin{align}\label{FIMHAD11}
		\mathbf{F}_h = \frac{{8{\pi ^2}\mathbf{SNR}^2}{{\cos }^2}{\theta _0}}{{{\lambda ^2}{M_h}\Xi }}\left[ {\frac{1}{{12}}{{\left\| {{b_h}({\theta _0})} \right\|}^4}M_h^2K_h^2\left( {K_h^2 - 1} \right){d^2} + \frac{{{M_h}{K_h}}}{\Xi }\left( {{{\left\| {{b_h}({\theta _0})\mu } \right\|}^2} + {K_h}\Re\left\{ {{b_h}^2({\theta _0})\mu } \right\}} \right)} \right],
	\end{align}
	\hrulefill
	\vspace*{4pt}
\end{figure*}
\begin{align}
	\mu = \sum_{m=1}^{M_h}\left(m-1\right)d e^{-j\frac{2\pi}{\lambda}(m-1)d\sin{\theta _0}}
\end{align}
\begin{align}
	\Xi  = {M_h} + {K_h}\mathbf{SNR}{\left\| {{b_h}({\theta _0})} \right\|^2}
\end{align}

The closed-form expression of the $C R L B_{h}$ is expressed by (\ref{CRLB2}).
\begin{figure*}[t]
	\begin{align}\label{CRLB2}
		CRLB_{h} &= \frac{1}{N}\mathbf{F}_{h}^{-1}= =\frac{\lambda ^2{M_h}\Xi}{{8L{\pi ^2}\mathbf{SNR}{{\cos }^2}{\theta _0}}\left[{\frac{{{\left\| {{b_h}({\theta _0})} \right\|}^4}M_h^2 K_h^2\left( {K_h^2 - 1} \right){d^2}}{{12\Xi}} + \frac{{M_h{K_h}}}{\Xi}\left( {{{\left\| {{b_h}({\theta _0})\mu } \right\|}^2} + {K_h}\Re\left\{ {{b_h}^2({\theta _0})\mu } \right\}} \right) }
 \right]}
	\end{align}
	\hrulefill
	\vspace*{4pt}
\end{figure*}

\textit{Theorem 1:} The closed-form expression of $w_{FD}$ and $w_{h}$ is respectively expressed by
\begin{align} \label{wqFinal1}
	w_{FD}=\frac{C R L B_{FD}^{-1}}{C R L B_{FD}^{-1}+\sum_{h=1}^{H}
 C R L B_{h}^{-1}}
\end{align}
\begin{align} \label{wqFinal2}
 w_h=\frac{C R L B_{h}^{-1}}{C R L B_{FD}^{-1}+\sum_{h=1}^{H}
 C R L B_{h}^{-1}},
\end{align}
\textit{Proof:} See Appendix A.

Based on the above analysis, the proposed CoMD framework contains three steps: 1) form the candidate sets for $\rm{H}^2AD$ and initial coarse DOA value for FD subarray, respectively. 2) infer the true solution for each group by transferring the coarse DOA estimation
$\hat{\theta}_{FD}$ to $\rm{H}^2AD$. 3) fuse two-part true solution. The overall algorithm is described in Algorithm \ref{alg:GS}.

\begin{algorithm}[t]
	\caption{Proposed CoMD.}\label{alg:GS}
	\begin{algorithmic}
		\STATE 
		\STATE {\textbf{Input:}}$~\mathbf{y}(n),~ n=1,2,\cdots,L.$
		\STATE \hspace{0.5cm} \textbf{Initialization:} ~divide $\mathbf{y}(n)$ into $\mathbf{y}_{FD}(n)$ and $\mathbf{y}_h(n)$, $h=1,2,\cdots,H$. Calculate the $CRLB_{h}$ and $CRLB_{FD}$.
		\STATE \hspace{0.5cm} \textbf{for} $h=1,2,\cdots,H$ \textbf{do},
		\STATE \hspace{1cm} perform the root-MUSIC method for $\mathbf{y}_h(n)$, and obtain the solution set, $\hat{\Theta}_{h}$.
		\STATE \hspace{0.5cm} \textbf{end for}
		\STATE \hspace{0.5cm} treat $H$ solutions as $H$ clusters
		\STATE \hspace{0.5cm} \textbf{repeat}
            \STATE \hspace{0.5cm} Calculate the $\hat{\theta}_{FD}$ by the root-MUSIC or CNN-based method for $\mathbf{y}_{FD}(n)$.
            \STATE \hspace{0.5cm} Solve the problem (\ref{w_op}) to obtain the set $\hat{\Theta}_{t}'$.
            \STATE \hspace{0.5cm} Integrate the $\hat{\Theta}_{t}'$,(\ref{wqFinal1}) and (\ref{wqFinal2}) into (\ref{thetaEstConbine}) to get $\hat{\theta}$.
		\STATE {\textbf{Output:}} $\hat{\theta}$
	\end{algorithmic}
\end{algorithm}


\section{Theoretical Analysis}\label{sec_perf}
This section provides the analysis of the theoretical characterization for the $\rm{H}^2$AD structure and the complexity of the proposed approaches.
\subsection{CRLB}
The CRLB provides a lower bound of the variance for an unbiased DOA method. Therefore, the deriving process of CRLB for the proposed FD$\rm{H}^2$AD structure is described in theorem 1.

\textit{Theorem 2:} The CRLB for the proposed FD$\rm{H}^2$AD structure is expressed by
\begin{align} \label{FFFFSSSS}
	\sigma_{\theta_0}^2 \geq \frac{1}{L}\mathbf{F}_\mathbf{y}^{-1}
\end{align}
where
\begin{align} \label{}
	\mathbf{F}_\mathbf{y} =\mathbf{F}_{FD}+\mathbf{F}_{\mathbf{\rm{H}^2}AD}=\mathbf{F}_{FD}+\sum_{h=1}^{H}\mathbf{F}_{h}
\end{align}
\textit{Proof:} See Appendix B.

\subsection{MSE}
This subsection presents the MSE of the proposed multi-modal-learning-based
framework. Assuming that the DOA estimation of FD and $\rm{H}^2$AD array can realize the CRLB. Also, we could obtain all correct solutions. According to (\ref{MAEA})-(\ref{wqFinal}), the MSE of the proposed framework is given by 

\begin{equation} 
\begin{aligned} \label{MSEaaA}
	&\mathbf{MSE}= w_{FD}^2 CRLB_{FD}+\sum_{h=1}^{H} w_{h}^2CRLB_{h}\\
 &=\left(\frac{C R L B_{FD}^{-1}}{C R L B_{FD}^{-1}+\sum_{h=1}^{H}C R L B_{h}^{-1}}\right)^2 CRLB_{FD}+ \\
 &\quad\sum_{h=1}^{H}\left( \frac{C R L B_{h}^{-1}}{C R L B_{FD}^{-1}+\quad\quad\sum_{h=1}^{H}C R L B_{h}^{-1}}\right)^2 CRLB_{h}
 \\
 &=\frac{1}{C R L B_{FD}^{-1}+\sum_{h=1}^{H}C R L B_{h}^{-1}}
\end{aligned}
\end{equation}
where  

\begin{equation}
	CRLB_{FD} =\frac{1}{L\cdot\mathbf{F}_{FD}}
\end{equation}
\begin{equation}
  CRLB_{h}=\frac{1}{L\cdot\mathbf{F}_h},
\end{equation}
then 
\begin{equation}
	\mathbf{MSE} =\frac{1}{L}\left(\mathbf{F}_{FD}+\sum_{h=1}^{H}\mathbf{F}_{h}\right)^{-1}
  =\frac{1}{L}\left(\mathbf{F}_{FD}+\mathbf{F}_{\mathbf{\rm{H}^2}AD}\right)^{-1},
\end{equation}
Obviously, the MSE is identical to (\ref{FFFFSSSS}). Hence, the
multi-modal-learning-based framework is consistent with the corresponding CRLB.

%

\section{Simulation Results}\label{sec_simu}

In this section, experimental results are presented to assess the performance of our proposed DOA estimator and the vital parameters of simulation settings will be given in Table \ref{tableDOA}. Furthermore, we use the root-mean-squared error (RMSE) to represent the performance and the hybrid CRLB as a baseline, which is calculated as

\begin{align} \label{}
	RMSE = \sqrt {\frac{1}{U}\sum_{u}^{U}(\hat{\theta}_{u}-\theta_0)^2}
\end{align}
where $U$ denotes the number of Monte Carlo experiments.

\begin{table}[http]
\caption{SYSTEM DATA SET PARAMETERS}
\label{tableDOA}
\centering
\begin{tabular}{ccl}
\cline{1-2}
Parameters  & Values &  \\ 
\cline{1-2}
Number of groups in $\rm{H}^2$AD: $H$ & 3 &\\
Number of subarrays in each group: $K_1, K_2, K_3$ & 16, 16, 16 &\\
Number of antennas in each group: $M_1, M_2, M_3$     & 7, 11, 13 &  \\
Number of antennas of FD array & 128 &  \\
Transmitter direction: $\theta_0$ & 41$^\circ$  &  \\
Monte Carlo experiments: U & 5000 &  \\
Number of snapshots:: $L$& 100 &\\
Conv\_1: kernel size, stride  & $3 \times 3@256$, 2 &  \\
Conv\_2-Conv\_5: kernel size, stride  & $2 \times 2@256$, 1 &  \\
Neurons in FC\_1, FC\_2, FC\_3 & 1024, 512, 256 &  \\
\cline{1-2}
\end{tabular}
\end{table}

\begin{figure}[!htb]
	\centering	\includegraphics[width=3.5in]{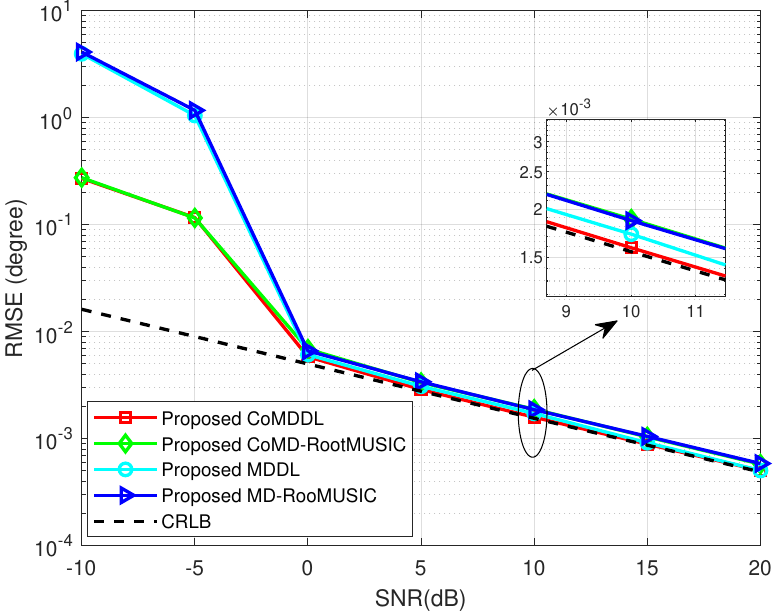}
	\caption{RMSE versus SNR of the proposed method.}\label{RMSE_SNR}
\end{figure}

Figure.\ref{RMSE_SNR} illustrates the 
curves of our proposed DOA methods in RMSE versus SNR, with the corresponding CRLB as the performance baseline. From Figure.\ref{RMSE_SNR}, 
the proposed four methods could approach the corresponding CRLB when SNR > 0 dB.
In particular, as SNR tends to the extremely SNR region, the performance gains over MDDL and MD-RootMUSIC achieved by CoMDDL and CoMD-RootMUSIC grow gradually. At SNR=-10dB, the learned precision of the latter is ten times the former.

\begin{figure}[!htb]
	\centering	\includegraphics[width=3.5in]{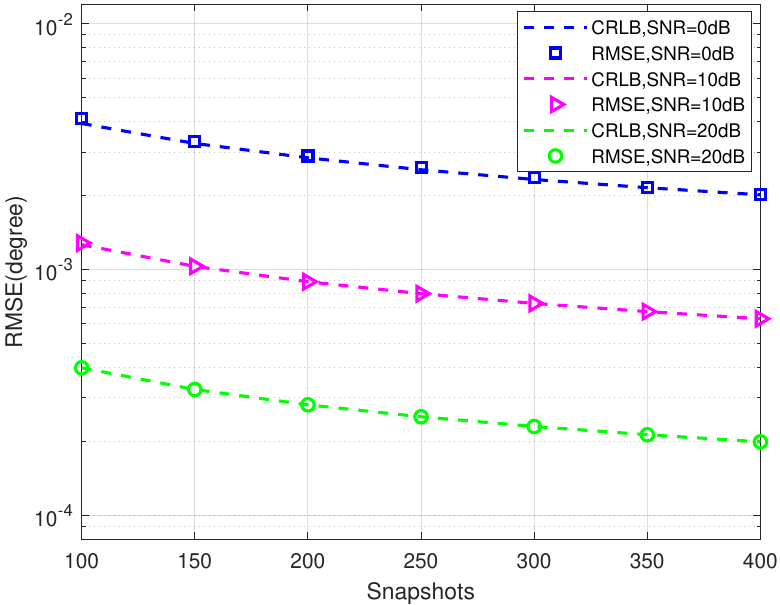}
	\caption{RMSE versus number of snapshots with SNR$ \in \{0, 10, 20\}$ dB.}\label{snapshot}
\end{figure}

Figure.\ref{snapshot} plots the curve of RMSE versus the number of snapshots $L$ under 
SNR$\in\{0,10,20\}$dB. 
From Figure.\ref{snapshot}, as the number of snapshots $L$ increases, the RMSE performance gradually improves.
Moreover, at SNR$ = 0$, the performance can approach the corresponding CRLB when $L$ reaches 150.
Meanwhile, the desired DOA estimation can be achieved at low-number snapshots as an increase in SNR.

\begin{figure}[!htb]
	\centering	\includegraphics[width=3.4in]{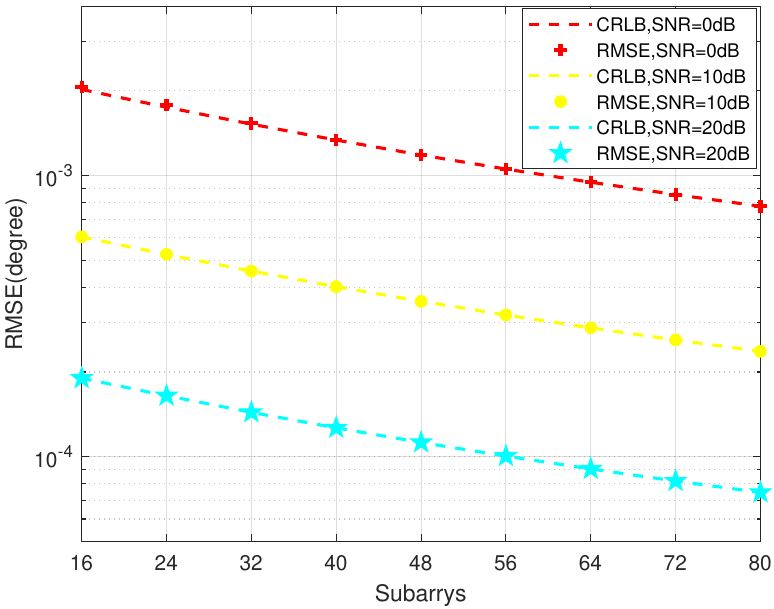}
	\caption{RMSE versus number of subarrys with SNR$ \in \{0, 10, 20\}$ dB.}\label{subarry}
\end{figure}

Figure.\ref{subarry} depicts RMSE versus the number of subarrays, where $M1=7$, $M2=11$, $M3=13$, $K1 = K2 =K3$ $(\in\{16:8:80\}$. It can be observed from the figure that the corresponding CRLB at SNR$\in\{0,10,20\}$dB can be realized at almost all points of RMSE as $K_h$ increases. Combining the experimental results in Figure.\ref{RMSE_SNR} and Figure.\ref{snapshot} , the proposed DOA estimator can realize a superior performance in medium and high SNR.

\begin{figure}[!htb]
	\centering	\includegraphics[width=3.5in]{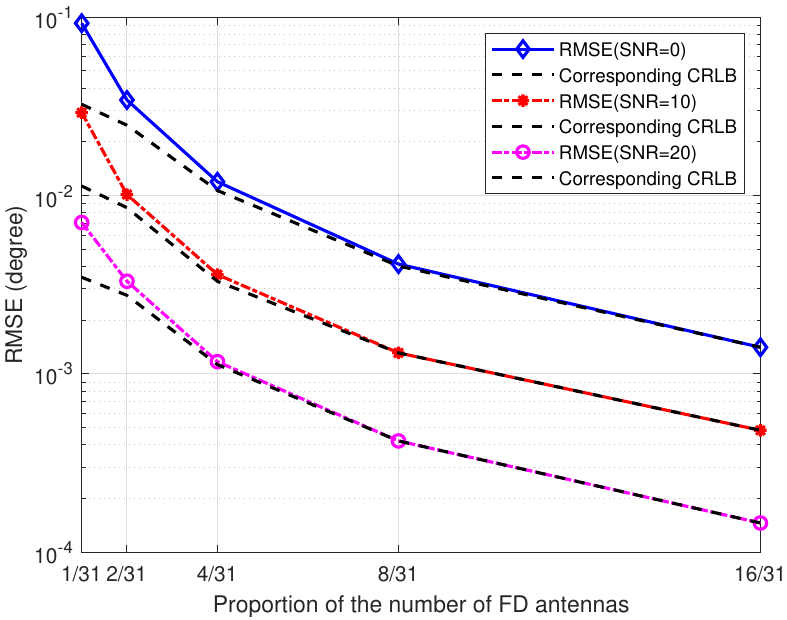}
	\caption{RMSE versus proportion of FD.}\label{Proportion}
\end{figure}

Figure.\ref{Proportion} plots the curves between RMSE and the ratio of FD structure antennas to the total antenna under SNR$\in\{0,10,20\}$dB. From Figure.\ref{Proportion}, 
the DOA estimation performance gradually improves as the proportion of the FD antennas increases, and the RMSE performance can reach the CRLB 
when the proportion is up to $4/31$. This implies different scenes should choose the suitable FD antennas to achieve DOA estimation.

\section{Conclusions}\label{sec_con}

In this paper, a novel universal $\rm{H}^2AD$-FD MIMO receiver structure of integrating FD subarray with heterogeneous HAD structure is designed. It effectively solves the issue of high latency or low time-efficiency in DOA estimation encountered with conventional HAD structure.
The primary advantage of this structure lies in its ability to rapidly eliminate phase ambiguity within a single time-slot, thereby enhancing the time-efficiency of DOA measurements. 
Based on this structure, two three-step MDL frameworks, MD and CoMD, for measuring DOA without ambiguity is proposed. The CoMD framework establishes a relation between FD and $\rm{H}^2AD$ modals by leveraging a prior knowledge of FD, thereby further improving DOA performance and reducing the clustering complexity. Moreover, CNN designed to be a five-layer neural network can obtain more accurate starting sample point of
true solution class. 
Simulation and analysis results demonstrate that the proposed four methods could approach the corresponding CRLB. In particular, the proposed CoMDDL and MDDL exhibit more excellent DOA performance than  MD-RootMUSIC and CoMD-RootMUSIC in the extremely low-SNR scenarios.
Furthermore, the structure, frameworks and methodologies presented in this paper strike a favorable balance in terms of circuit cost, time-efficiency, phase ambiguity elimination and accuracy.
Leveraging these advantages, the proposed $\rm{H}^2AD$-FD-based DOA measurements hold promise for the development of green wireless communication technologies in the future.

\section*{APPENDIX A: Proof of Theorem 1}
Based on (\ref{w_op}) and the Lagrangian theorem, we can define $L(w_{FD},w_h,\zeta)$, as
\begin{equation} 
\begin{aligned} \label{lagFunc}
 &L(w_{FD},w_h,\zeta)=\sum_{h=1}^{H} w_{h}^2CRLB_{h}+w_{FD}^2 CRLB_{FD}\\
 &\quad\quad\quad\quad\quad\quad-\zeta(w_{FD} + \sum_{h=1}^{H} w_{h}-1)
\end{aligned}
\end{equation}
The partial derivatives of (\ref{lagFunc}) are expressed as 
\begin{equation}
	\frac{\partial L(w_{FD},w_h,\zeta)}{\partial w_{FD}}= 2 w_{FD} CRLB_{FD}-\zeta
\end{equation}
\begin{equation}
	\frac{\partial L(w_{FD},w_h,\zeta)}{\partial w_h}= 2 w_h CRLB_h-\zeta
\end{equation}
\begin{equation}
	\frac{\partial L(w_{FD},w_h,\zeta)}{\partial \zeta}=- w_{FD}-\sum_{h=1}^{H}w_h+1
\end{equation}
Then, we can set
\begin{equation}
		\begin{split}
		\left \{
		\begin{array}{ll}
			\frac{\partial L(w_{FD},w_h,\zeta)}{\partial w_{FD}} = 0\\
			\frac{\partial L(w_{FD},w_h,\zeta)}{\partial w_h}= 0\\
                \frac{\partial L(w_{FD},w_h,\zeta)}{\partial \zeta}=0
		\end{array}
		\right.
	\end{split}
\end{equation}
which yields
\begin{align} \label{wqFinal}
	&w_{FD}=\frac{C R L B_{FD}^{-1}}{C R L B_{FD}^{-1}+\sum_{h=1}^{H}
 C R L B_{h}^{-1}}\nonumber\\
 &w_h=\frac{C R L B_{h}^{-1}}{C R L B_{FD}^{-1}+\sum_{h=1}^{H}
 C R L B_{h}^{-1}},
\end{align}
Therefore, the proof of Theorem 1 is completed.

\section*{APPENDIX B: CRLB for the $\rm{H}^2$AD-FD structure}\label{AppB}
In this section, the CRLB for the $\rm{H}^2$AD-FD structure is derived. According to \cite{shuHADDOA2018tcom}, the Fisher information matrix for $\rm{H}^2$AD-FD structure can be given by
\begin{equation}\label{fimm}
	\mathbf{F}_{\mathbf{y}} = \mathbf{Tr}\left\{
	\mathbf{R}_{\mathbf{y}}^{-1} \frac{\partial \mathbf{R}_{\mathbf{y}}}{\partial \theta}\mathbf{R}_{\mathbf{y}}^{-1} \frac{\partial \mathbf{R}_{\mathbf{y}}}{\partial \theta}
	\right\}
\end{equation}
where
\begin{align}\label{yyyyy}
    \mathbf{y} & = \left[
    \begin{array}{l}
    {\mathbf{y}_{FD}}\\
    {\mathbf{y}_{{\rm{H}^2}AD}}
    \end{array} 
\right] \mathop  = \limits^\Delta  \mathbf{G}_A^Has + \mathbf{w} \nonumber\\
& =\left[ 
\begin{array}{*{20}{c}}
1&0\\
0&{\mathbf{\Upsilon} _A^H}
\end{array}
\right]\left[ 
\begin{array}{l}
{\mathbf{a}_{FD}}\\
{\mathbf{a}_{{\rm{H}^2}AD}}
\end{array} \right]s + \left[ \begin{array}{l}
{\mathbf{w}_1}\\
{\mathbf{w}_2}
\end{array} 
\right]
\end{align}

\begin{equation}\label{RTy}
	\mathbf{R}_{\mathbf{y}} = \mathbf{SNR} \mathbf{G}_A^H \mathbf{a}\mathbf{a}^H \mathbf{G}_A + \mathbf{I}
\end{equation}
Therefore, $\mathbf{R}_{\mathbf{y}}$ can be expressed by (\ref{RysuBb})
\begin{figure*}[ht]
\begin{align}\label{RysuBb}
		\mathbf{R}_{\mathbf{y}} &=\mathbf{SNR}
  \left[ 
\begin{array}{*{20}{c}}
1&0\\
\mathbf{0}&{\mathbf{\Upsilon} _A^H}
\end{array}
\right]\left[ 
\begin{array}{l}
{\mathbf{a}_{FD}}\\
{\mathbf{a}_{{\rm{H}^2}AD}}
\end{array} \right]
\left[ {\mathbf{a}_{FD}^H,\mathbf{a}_{{\rm{H}^2}AD}^H} \right]
\left[\begin{array}{*{20}{c}}
1&0\\
\mathbf{0}&{\mathbf{\Upsilon} _A}
\end{array}
\right]+\mathbf{I}	\nonumber\\
& = \mathbf{SNR}
  \left[ 
\begin{array}{*{20}{c}}
{\mathbf{a}_{FD}}{\mathbf{a}_{FD}^H}&\mathbf{0}\\
\mathbf{0}&{\mathbf{\Upsilon} _A^H}{\mathbf{a}_{{\rm{H}^2}AD}}{\mathbf{a}_{{\rm{H}^2}AD}^H}{\mathbf{\Upsilon} _A}
\end{array}
\right]+\mathbf{I}
=\left[
\begin{array}{*{20}{c}}
{\mathbf{R}_{FD}}&\mathbf{0}\\
\mathbf{0}&{\mathbf{R}_{{\rm{H}^2}AD}}
\end{array}
\right]
\end{align}
\hrulefill
\vspace*{4pt}
\end{figure*}

Then, $\mathbf{F}_{\mathbf{y}}$ is given by (\ref{FIMQQ}),
\begin{figure*}[ht]
	\begin{align}\label{FIMQQ}
		\mathbf{F}_{\mathbf{y}} &=\mathbf{Tr}\left\{ \left[
            {\begin{array}{*{20}{c}}
            \mathbf{R}_{FD}^{-1} \frac{\partial \mathbf{R}_{{FD}}}{\partial \theta}\mathbf{R}_{FD}^{-1} \frac{\partial \mathbf{R}_{FD}}{\partial \theta}&\mathbf{0}\\
            \mathbf{0}&{\mathbf{R}_{{\rm{H}^2}AD}^{-1} \frac{\partial \mathbf{R}_{{\rm{H}^2}AD}}{\partial \theta}\mathbf{R}_{{\rm{H}^2}AD}^{-1} \frac{\partial \mathbf{R}_{{\rm{H}^2}AD}}{\partial \theta}}
            \end{array}}
		\right]
		\right\} = \mathbf{F}_{FD}+\mathbf{F}_{{\rm{H}^2}AD}
	\end{align}	
	\hrulefill
	\vspace*{4pt}
\end{figure*}
where $\mathbf{F}_{FD}$ is
\begin{align}\label{FIMFD}
\mathbf{F}_{FD} = \frac{{8{\pi ^2}\mathbf{SNR}{{\cos }^2}{\theta _0}\mathop {{d^2}}\limits^ - }}{{\lambda ^2}}
\end{align}

And the deriving process of $\mathbf{F}_{{\rm{H}^2}AD}$ as follows

\begin{align}\label{yyyy}
\mathbf{y}_{{\rm{H}^2}AD} &=\left[\mathbf{y}_1,\mathbf{y}_2,\cdots,\mathbf{y}_H\right]^T 
=  \mathbf{\Upsilon}_A^H \mathbf{a_{{\rm{H}^2}AD}}s+\mathbf{w}\nonumber\\
 & \mathop  = \limits^\Delta  
 \left[
	\begin{array}{cccc}
		\mathbf{\Upsilon}_{A,1}^H & \mathbf{0} & \cdots & \mathbf{0} \\
		\mathbf{0} & \mathbf{\Upsilon}_{A,2}^H & \cdots & \mathbf{0} \\
		\vdots & \vdots & \ddots & \vdots \\
		\mathbf{0} & \mathbf{0} & \cdots & \mathbf{\Upsilon}_{A,H}^H
	\end{array}
	\right]\mathbf{a}_{{\rm{H}^2}AD}s+  \mathbf{w} 
\end{align}
where
\begin{equation}
	\mathbf{a}_{{\rm{H}^2}AD} = \left[\tau_1\mathbf{a}_1^T,\tau_2\mathbf{a}_2^T,\cdots,\tau_H\mathbf{a}_H^T\right]^T
\end{equation}
And $\tau_h$ is the phase of the $h$th array corresponding to the leftmost antenna
\begin{equation}\label{varphiq}
	\tau_h = \left\{
	\begin{array}{ll}
		1 &,h=1 \\
		e^{j \frac{2 \pi}{\lambda} d \sin \theta_0\sum_{h_j=1}^{h-1}N_{h_j} }  &,h>1
	\end{array}
	\right.
\end{equation}
Thus, $\mathbf{R_{{\rm{H}^2}AD}}$ is given by (\ref{RyH2AD})
\begin{figure*}[ht]
	\begin{align}\label{RyH2AD}
		\mathbf{R}_{\mathbf{\rm{H}^2}AD} &=\mathbf{SNR}
		\left[
		\begin{array}{cccc}
			\mathbf{\Upsilon}_{A,1}^H & \mathbf{0} & \cdots & \mathbf{0} \\
			\mathbf{0} & \mathbf{\Upsilon}_{A,2}^H & \cdots & \mathbf{0} \\
			\vdots & \vdots & \ddots & \vdots \\
			\mathbf{0} & \mathbf{0} & \cdots & \mathbf{\Upsilon}_{A,H}^H
		\end{array}
		\right]\left[
		\begin{array}{c}
			\tau_1\mathbf{a}_1  \\
			\tau_2\mathbf{a}_2  \\
			\vdots  \\
			\tau_H\mathbf{a}_H
		\end{array}
		\right]\left[\tau_1^H\mathbf{a}_1^H, \cdots,\tau_H^H\mathbf{a}_H^H\right]
		\left[
		\begin{array}{cccc}
			\mathbf{\Upsilon}_{A,1} & \mathbf{0} & \cdots & \mathbf{0} \\
			\mathbf{0} & \mathbf{\Upsilon}_{A,2} & \cdots & \mathbf{0} \\
			\vdots & \vdots & \ddots & \vdots \\
			\mathbf{0} & \mathbf{0} & \cdots & \mathbf{\Upsilon}_{A,H}
		\end{array}
		\right]+\mathbf{I} \nonumber\\		
  &=\mathbf{SNR} \mathbf{diag}\left\{\left[\mathbf{\Upsilon}_{A,1}^H\mathbf{a}_1\mathbf{a}_1^H\mathbf{\Upsilon}_{A,1},\mathbf{\Upsilon}_{A,2}^H\mathbf{a}_2\mathbf{a}_2^H\mathbf{\Upsilon}_{A,2},\cdots,\mathbf{\Upsilon}_{A,H}^H\mathbf{a}_H\mathbf{a}_H^H\mathbf{\Upsilon}_{A,H}
\right]\right\}+\mathbf{I}\nonumber\\
        &=\mathbf{SNR}\mathbf{diag}\left\{\left[\mathbf{R}_{\mathbf{y}_1},\mathbf{R}_{\mathbf{y}_2},\cdots,\mathbf{R}_{\mathbf{y}_H}\right]\right\}
	\end{align}
	\hrulefill
	\vspace*{4pt}
\end{figure*}

According to (\ref{fimm}) and (\ref{RyH2AD}), $\mathbf{F}_{{\rm{H}^2}AD}$ is written as (\ref{fH2AD}), 
\begin{figure*}[ht]
	\begin{align}\label{fH2AD}
		\mathbf{F}_{\mathbf{\rm{H}^2}AD}= &=\mathbf{Tr}\left\{\mathbf{diag} \left(\left[\mathbf{R}_{\mathbf{y}_1}^{-1} \frac{\partial \mathbf{R}_{\mathbf{y}_1}}{\partial \theta}\mathbf{R}_{\mathbf{y}_1}^{-1} \frac{\partial \mathbf{R}_{\mathbf{y}_1}}{\partial \theta},\mathbf{R}_{\mathbf{y}_2}^{-1} \frac{\partial \mathbf{R}_{\mathbf{y}_2}}{\partial \theta}\mathbf{R}_{\mathbf{y}_2}^{-1} \frac{\partial \mathbf{R}_{\mathbf{y}_2}}{\partial \theta},\cdots,\mathbf{R}_{\mathbf{y}_H}^{-1} \frac{\partial \mathbf{R}_{\mathbf{y}_H}}{\partial \theta}\mathbf{R}_{\mathbf{y}_H}^{-1} \frac{\partial \mathbf{R}_{\mathbf{y}_H}}{\partial \theta}
		\right]\right)
		\right\} = \sum_{h=1}^{H}\mathbf{F}_{h}
	\end{align}	
	\hrulefill
	\vspace*{4pt}
\end{figure*}
where $\mathbf{F}_{h}$ in accordance with \cite{shuHADDOA2018tcom} expressed by 
\begin{figure*}[t]
	\begin{align}\label{FIMHAD1}
		\mathbf{F}_h = \frac{{8{\pi ^2}\mathbf{SNR}^2}{{\cos }^2}{\theta _0}}{{{\lambda ^2}{M_h}\Xi }}\left[ {\frac{1}{{12}}{{\left\| {{b_h}({\theta _0})} \right\|}^4}M_h^2K_h^2\left( {K_h^2 - 1} \right){d^2} + \frac{{{M_h}{K_h}}}{\Xi }\left( {{{\left\| {{b_h}({\theta _0})\mu } \right\|}^2} + {K_h}\Re\left\{ {{b_h}^2({\theta _0})\mu } \right\}} \right)} \right],
	\end{align}
	\hrulefill
	\vspace*{4pt}
\end{figure*}

The CRLB can be given by
\begin{align}\label{CRLBH2AM}
	CRLB = \frac{1}{L}\mathbf{F}_{\mathbf{y}}^{-1}
 =\frac{1}{L}\left(\mathbf{F}_{FD}+\mathbf{F}_{\mathbf{\rm{H}^2}AD}\right)^{-1}
\end{align}

Then, submitting (\ref{FIMFD}) and (\ref{fH2AD}) into (\ref{CRLBH2AM}), we can obtain the closed-form expression of the CRLB.
\begin{figure*}[t]
	\begin{align}\label{FIMHAD1}
		CRLB = \frac{\lambda ^2}{{8L{\pi ^2}\mathbf{SNR}{{\cos }^2}{\theta _0}}\left[\mathop {{d^2}}\limits^ - +\sum_{h=1}^{H}\left({\frac{{{\left\| {{b_h}({\theta _0})} \right\|}^4}M_h K_h^2\left( {K_h^2 - 1} \right){d^2}}{{12\Xi}} + \frac{{{K_h}}}{\Xi^2 }\left( {{{\left\| {{b_h}({\theta _0})\mu } \right\|}^2} + {K_h}\Re\left\{ {{b_h}^2({\theta _0})\mu } \right\}} \right) }
 \right)\right]}
	\end{align}
	\hrulefill
	\vspace*{4pt}
\end{figure*}

The derivation of CRLB for proposed FD$\rm{H}^2$AD structure is completed.

\section*{Data Availability}
The data supporting the conclusions of this article are included in the article.

\section*{Conflicts of Interest}
The authors declare that they have no conficts of interest.


\bibliographystyle{IEEEtran}
\bibliography{reference}

\vfill\pagebreak

\end{document}